\documentclass{rnaastex}

\newcommand\name{$\!$`Oumuamua}
\begin{document}


\title{Kinematics of the Interstellar Vagabond
  1I/$\!$`Oumuamua (A/2017 U1)}

\correspondingauthor{Eric Mamajek}
\email{mamajek@jpl.nasa.gov}

\author[0000-0003-2008-1488]{Eric Mamajek}
\affiliation{Jet Propulsion Laboratory, California Institute of Technology, M/S
  321-100, 4800 Oak Grove Drive, Pasadena, CA 91109, USA}
\affiliation{Department of Physics \& Astronomy, University of
  Rochester, Rochester, NY 14627, USA}


\section{}

The discovery of an asteroid of likely interstellar origin was
recently made by the Pan-STARRS survey -- A/2017 U1 = 1I/\name
\footnote{See: http://www.minorplanetcenter.net/mpec/K17/K17UI1.html,
  https://www.minorplanetcenter.net/mpec/K17/K17V17.html.}.  {\it Can
 \name's velocity before it entered the solar system provide any
  clues to its origin?}
The best available orbit from the JPL Small-Body Database
Browser\footnote{https://ssd.jpl.nasa.gov/sbdb.cgi?sstr=A\%2F2017\%20U1}
(solution JPL-13 produced by Davide Farnocchia) lists perihelion
distance $q$ = 0.255287\, $\pm$\, 0.000079 au,
eccentricity $e$ = 1.19936\, $\pm$\, 0.00021 
and semi-major axis $a$ = -1.28052\, $\pm$\, 0.00096 au.
%
%
%
This value of $a$ is consistent with an initial velocity before
encountering the solar system of $v_{\circ}$ = 26.3209\,$\pm$\,0.0099
km\,s$^{-1}$, assuming no non-gravitational forces.
The ephemeris shows that the object entered the solar system from the
direction $\alpha_{ICRS}$, $\delta_{ICRS}$ = 279$^{\circ}$.804,
+33$^{\circ}$.997 ($\pm$0$^{\circ}$.032, $\pm$0$^{\circ}$.015;
1$\sigma$).
This divergent point and $v_{\circ}$ value translates to a
heliocentric Galactic velocity \citep[][$U$ towards Galactic
  center]{Perryman98} of $U, V, W$ = -11.457, -22.395, -7.746
km\,s$^{-1}$ ($\pm$0.009, $\pm$0.009, $\pm$0.011 km\,s$^{-1}$).\\

%

{\it Could\, \name\, be a member of the Oort Cloud of the $\alpha$
  Centauri system?}
Such a scenario might not be unexpected as the tidal radius
for the 2.17 $M^{N}_{\odot}$ triple system \citep{Kervella17} is of
order $r_t$ $\simeq$ 1.7 pc \citep[][]{Mamajek13}.
As the system lies only 1.34 pc away, the solar system may be on the
outskirts of $\alpha$ Cen's cometary cloud \citep[see][]{Hills81,Beech11}.
\citet{Kervella17} calculated updated heliocentric Galactic
velocities for $\alpha$ Cen AB of $U, V, W$ = -29.291, 1.710, 13.589
($\pm$0.026, $\pm$0.020, $\pm$0.013) km\,s$^{-1}$ and for Proxima
Centauri ($\alpha$ Cen C) of $U, V, W$ = -29.390, 1.883, 13.777
($\pm$0.027, $\pm$0.018, $\pm$0.009) km\,s$^{-1}$.
The velocity difference of 36.80\,$\pm$\,0.04 km\,s$^{-1}$ between
\name\, and the $\alpha$ Cen system, and the fact they were further
apart in the past ($\Delta$ $\simeq$ 5 pc 100 kyr ago), {\it argues
  that it has no relation to $\alpha$ Cen}.
Members of $\alpha$ Cen's cometary cloud would appear to have motions
diverging from the vicinity of $\alpha$, $\delta$ = 293$^{\circ}$,
-42$^{\circ}$ with $v_{\circ}$ $\simeq$ 32 km\,s$^{-1}$.\\



\begin{figure}[h!]
\begin{center}
\includegraphics[scale=0.7,angle=0]{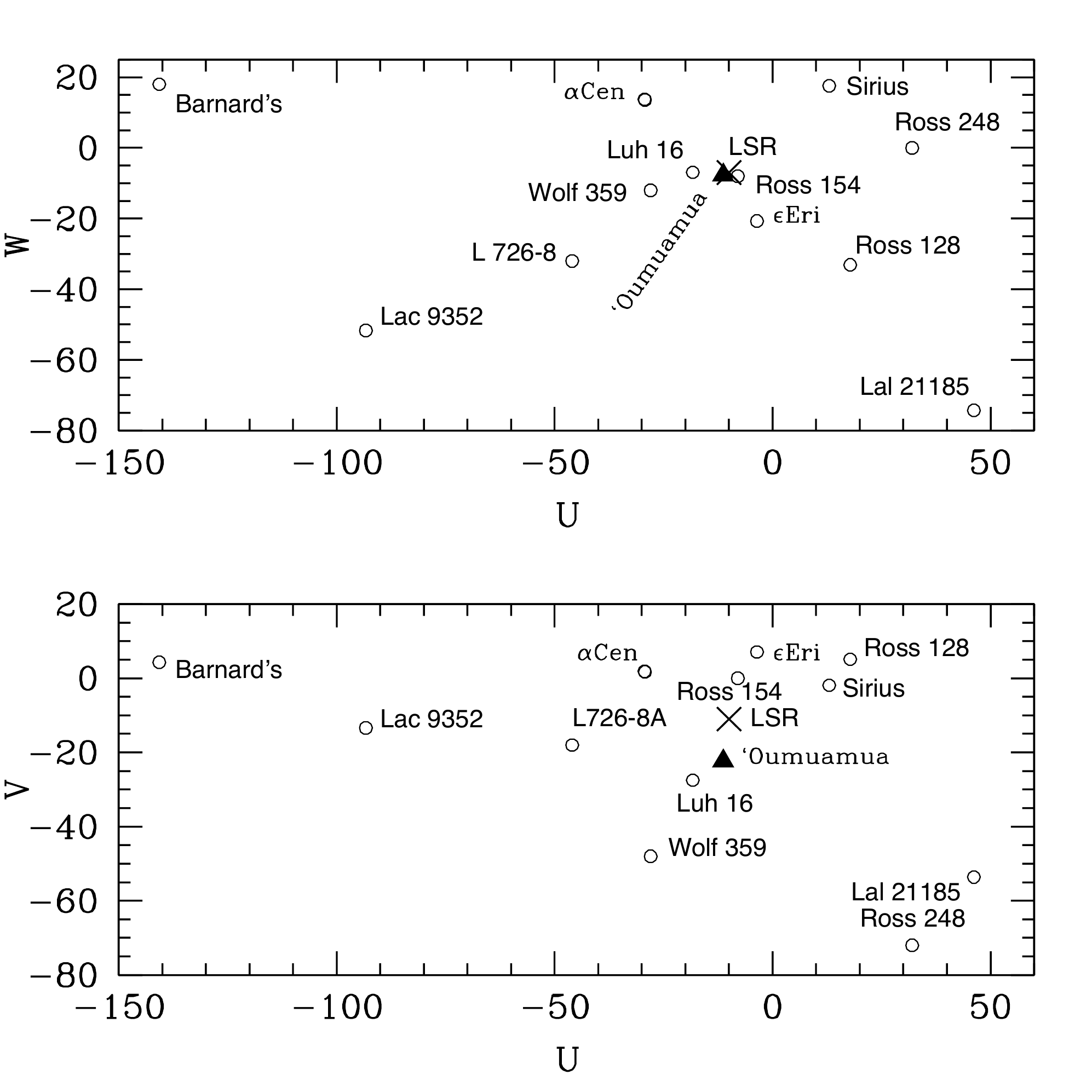}
\caption{Galactic velocities for 1I/\name\, (filled triangle),
  nearby stars (open circles), and LSR (cross)\label{fig1}.}
\end{center}
\end{figure}

The Galactic velocity of\, $\!$`Oumuamua is plotted against those of
the nearest stars (parallax $>$ 300 mas) in Fig.  \ref{fig1}.
Besides the velocity of $\alpha$ Cen AB and C from \citet{Kervella17},
velocities for the nearest stars are drawn from \citet{Anderson12} and
\citet{Hawley97}.
The velocity for the substellar binary Luhman 16 is calculated using
data from \citet{Garcia17} and \citet{Kniazev13}: $U, V, W$ = -18.3,
-27.5, -6.9 km\,s$^{-1}$.
\name's velocity is more than 20 km\,s$^{-1}$ from any of the stars,
and 9 km\,s$^{-1}$ off from Luhman 16, so {\it \name\, does not appear
  to be comoving with any of these nearest systems}.\\

{\it What velocities might be expected of interstellar field objects?}
We might first suspect that interstellar planetesimals share the
velocity distribution of nearby stars.
The XHIP catalog \citep{Anderson12} contains velocities for 1481 stars
within 25 pc with distances of $<$10\% accuracy.
The XHIP sample has median velocity $U, V, W$ = -10.5, -18.0, -8.4
km\,s$^{-1}$ ($\pm$33, $\pm$24, $\pm$17 km\,s$^{-1}$; 1$\sigma$
range), similar to that for volume-limited samples of nearby M dwarfs
\citep[$U, V, W$ = -9.7, -22.4, -8.9 km\,s$^{-1}$ ; $\pm$37.9,
  $\pm$26.1, $\pm$20.5; 1$\sigma$;][]{Reid02}.
\citet{BlandHawthorn16} provides a recent consensus estimate for the
Local Standard of Rest (LSR) of $U, V, W$ = -10.0, -11.0, -7.0
km\,s$^{-1}$ ; $\pm$1, $\pm$2, $\pm$0.5; 1$\sigma$).
 An object with the median velocity of the local XHIP sample would
 have speed 22.5 km\,s$^{-1}$ coming from $\alpha$, $\delta$ =
 273$^{\circ}$, +33$^{\circ}$, within only $\sim$6$^{\circ}$ of\, \name's
 divergent point.
The velocity is very close to the median for the XHIP sample ($\Delta
v$ $\simeq$ 4.5 km\,s$^{-1}$; $\chi^2$/$\nu$ = 0.036/3; P = 0.0018),
the mean for the local M dwarfs ($\Delta v$ $\simeq$ 2.1 km\,s$^{-1}$;
$\chi^2$/$\nu$ = 0.0053/3; P = 0.0001) and the LSR ($\Delta v$
$\simeq$ 11.5\,$\pm$\,2.3 km\,s$^{-1}$), compared to the typical 3D
velocity of nearby stars.  Compared to the LSR, \name\, has negligible
radial and vertical Galactic motion\footnote{\citet{Gaidos17} have
  proposed that the object's velocity is due to birth in a nearby
  $\sim$40 Myr stellar association.}, and its sub-Keplerian circular
velocity trails by 11 km\,s$^{-1}$.\\

Robotic reconnaissance of \name, or future interstellar planetesimals
passing through the solar system, might constitute logical precursors
to interstellar missions, and provide the opportunity to conduct
chemical studies and radiometric dating of extrasolar material which
formed around other stars.\\

\acknowledgments

This research was carried out at the Jet Propulsion Laboratory,
California Institute of Technology, under a contract with the National
Aeronautics and Space Administration.  EEM acknowledges support from
the NASA NExSS program, and thanks Davide Farnocchia (JPL) for
discussions.  This work used the JPL Small-Body Database Browser,
HORIZONS system, and Vizier.\\

\end{document}